\documentclass[superscriptaddress,amsmath,amssymb,aps,pre,onecolumn]{revtex4-1}
\usepackage{dcolumn}% Align table columns on decimal point
\usepackage{bm}% bold math
\usepackage{graphicx}
\usepackage{mathtools}
\usepackage{physics}
\usepackage{comment}
\usepackage{color}

\newcommand{\bea}{\begin{eqnarray}}

\newcommand{\eea}{\end{eqnarray}}

\newcommand{\blue}{\textcolor{black}}

\begin{document}

%\preprint{}
\title{Critical number of walkers for diffusive search processes with resetting}% Force line breaks with \\
\author{Marco Biroli}
\affiliation{LPTMS, CNRS, Univ.  Paris-Sud,  Universit\'e Paris-Saclay,  91405 Orsay,  France}
\author{Satya N. Majumdar}
\affiliation{LPTMS, CNRS, Univ.  Paris-Sud,  Universit\'e Paris-Saclay,  91405 Orsay,  France}
\author{Gr\'egory Schehr}
\affiliation{Sorbonne Universit\'e, Laboratoire de Physique Th\'eorique et Hautes Energies, CNRS UMR 7589, 4 Place Jussieu, 75252 Paris Cedex 05, France}

%\date{\today}

\begin{abstract}
We consider $N$ Brownian motions diffusing independently on a line,
starting at $x_0>0$, in the presence of an absorbing target at the origin.
The walkers undergo stochastic resetting under two protocols: (A)
each walker resets {\em independently} to $x_0$ with rate $r$ and (B)
all walkers reset {\em simultaneously} to $x_0$ with rate $r$.
We compute analytically the mean first-passage time to 
the origin and show that, as a function of $r$ and for fixed $x_0$,
it has a minimum at an optimal value $r^*>0$ as long as $N<N_c$.
Thus resetting is beneficial for the search for $N<N_c$. 
When $N>N_c$, the optimal value occurs at $r^*=0$ indicating that resetting
hinders search processes. Continuing our results
analytically to real $N$, we show that $N_c=7.3264773\ldots$
for protocol A and $N_c=6.3555864\ldots$ for protocol B, independently
of $x_0$.
Our theoretical predictions are verified in numerical Langevin simulations.
\end{abstract}

\maketitle

%\tableofcontents

\newpage

\section{Introduction}

Search processes are ubiquitous in nature and human behavior \cite{bell, 
adam,Metzler_14} with examples ranging from foraging animals \cite{bartumeus, 
viswanathan} to proteins trying to bind on DNA \cite{berg, coppey, 
ghosh, chowdhury}. In most of these examples there is an interest in 
optimizing the search process, i.e., minimizing the time taken to reach 
the target by varying some underlying parameters
of the dynamics. One preeminent family of efficient search processes are 
the so called {\it intermittent} search strategies \cite{benichou_05, benichou_07, 
benichou_11}. For these processes, the searcher or the agent alternates between 
short and long range steps. During the short-range steps the agent 
actively searches for the target. Instead, the long-range steps allow it 
to explore new areas of the space. Resetting search processes are  
examples of efficient intermittent search processes \cite{villen, luby, 
tong, lorenz}, where after a certain time, random or non-random, 
the agent gives up on its  
current path and restarts from some other place, for a recent review see 
\cite{evans_20}.

While the idea of \blue{introducing resetting in a search process} had been
used empirically before, a quantitative computation of the 
search time was performed first in Ref.~\cite{evans_11,evans_11_2}
in a simple model of a Brownian agent searching for a fixed target
in space. For example, in the simplest case in one dimension, consider
a fixed target at the origin and a Brownian searcher with diffusion
constant $D$ that starts at the initial position $x_0$ and resets
to $x_0$ after an exponentially distributed random time with
rate $r$. The target is found when the walker reaches the origin for
the first time at $t=t_f$. Hence the mean search time is
just the mean first-passage time (MFPT) $\langle t_f\rangle_r(x_0)$ to
the origin, starting from $x_0$. One of the main findings of 
Ref.~\cite{evans_11} was that while the MFPT diverges in the
absence of resetting ($r=0$), it is finite for $r>0$ and is
given by 
\bea
\langle t_f\rangle_r(x_0)= \frac{1}{r}\, \left(e^{\sqrt{r/D}\, 
|x_0|}-1\right)\, .
\label{em.1}
\eea
For a fixed $x_0$, the MFPT in Eq. (\ref{em.1}), as a function of $r$,
has a unique minimum at $r=r^*= R^*\, x_0^2/D$, where
the dimensionless optimal rate $R^*=2.53962\ldots$ is easily found 
by minimizing Eq. (\ref{em.1})
with respect to $r$, and is given by the unique
root of $\sqrt{R}-2+2\, e^{-\sqrt{R}}=0$. 
Thus, not only the resetting renders the
MFPT finite, it can even be optimized by choosing the resetting
rate to be $r^*$. Subsequently, 
numerous models of search processes with resetting
found the existence of an optimal 
$r^*$~\cite{EM_14,KMSS_14,MSS_15,PKE_16,Reu_16,MV_16,PR_17,BEM_17,CS_18,Bres_20,Pinsky_20,BMS_22}. 
For simple diffusion with resetting in one and two dimensions,
the optimal $r^*$ was measured recently in optical tweezer 
experiments~\cite{TPSRR_20,BBPMC_20,FBPCM_21}. 

One naturally wonders if resetting is always advantageous, i.e.,
whether the optimal $r^*$ is strictly positive. This question
has been addressed in several papers for general single
particle search process subject to resetting. It turns out that
in many search processes, the
optimal value of $r^*$ may undergo a transition from
a nonzero value (resetting is beneficial) to zero (resetting is detrimental),
as one tunes some additional parameter through a critical value in the 
underlying search process
~\cite{KMSS_14,CS15,CM15,Reu_16,PR_17,Bel_18,RMS19,ANBBD19,Pal_19,Pal_19_2,Vasquez_20,Vasquez_22,Vasquez_22_2}.
A simple example concerns the diffusive search of a fixed target at the origin
in one dimension as discussed above, but now the searcher, starting
and resetting to $x_0$, is confined
in a box $[-L,L]$ with reflecting boundary conditions~\cite{CS15}. 
As $L\to \infty$, the MFPT is given by Eq. (\ref{em.1}) with
a nonzero $r^*$. As $L$ decreases, the value of $r^*$ decreases
and for $L\le L_c$, \blue{the optimal resetting rate becomes zero, i.e, $r^*=0$~\cite{CS15}}. 
Treating $r^*$ as an order parameter of this resetting phase transition, some
models exhibit a first order transition (where $r^*$ drops abruptly to zero), 
while some others a continuous transition (with
$r^*$ vanishing continuously). In Ref.~\cite{Pal_19}, a Landau like
theory was developed to study this resetting phase transition with
$r^*$ as the order parameter.  

This issue of the existence of an optimal $r^*$ has not been addressed
so far, to the best of our knowledge, when the search for the target is
conducted by a team of $N$ searchers with stochastic
resetting. The purpose of this paper is to study the optimal $r^*$ as 
a function of $N$ in a simple model of $N$ diffusive searchers on a
line undergoing stochastic resetting at a constant rate $r$. To be
precise, we will consider
$N$ diffusing particles on a line each 
with diffusion constant $D$ and starting at the same initial position $x_0$,
with the target fixed at the origin. Since the search process is symmetric
with respect to the sign of $x_0$, we consider only $x_0>0$ without
any loss of generality.
For resetting, we will follow
two \blue{distinct} protocols. 

\begin{itemize}

\item {\bf Protocol A.} 
In this protocol, \blue{each one of the $N$ particles} diffuses and resets to $x_0$
{\it independently} with rate $r$~\cite{evans_11}.  
The positions of the particles are thus 
{\it uncorrelated} at all times. For a typical schematic representation of the
trajectories see Fig. 1a.   

\item {\bf Protocol B.} Here each \blue{one of the $N$ particles} diffuses independently,
but they \blue{all} reset {\it simultaneously} to $x_0$ with rate $r$~\cite{biroli_22}. 
This simultaneous resetting makes the particle positions {\it correlated}
at all times $t$. See Fig. 1b for typical trajectories under the
protocol B.

\end{itemize}

For $N=1$, the two protocols coincide, but they are different for $N>1$.
In protocol $A$, the particles remain noninteracting at all times.
This protocol was first studied in 
Ref.~\cite{evans_11} with the initial positions of the
searchers distributed uniformly
with density $\rho$ (i.e., $N\to \infty$ limit) on one
side of the target at the origin and the authors
computed exactly the survival probability of the target up to time $t$.
In a recent work~\cite{VAM22}, the two-time correlation function of the
maximum displacement of the $N$ particles (without a target) was studied
numerically. However, the MFPT to a target for fixed $N>1$ has not
been studied. \blue{Protocol B} was recently introduced in Ref.~\cite{biroli_22}
and it was shown that in the absence of a target, the
system approaches at long times a many-body 
nonequilibrium stationary state with strong correlations between the positions
of the particles. The stationary joint distribution of the 
positions of the particles was computed exactly. Despite strong correlations
between particles, several observables such as the distribution of
the position of the $k$-th
rightmost particle, the distribution of the successive gaps between particles
etc. were computed analytically in the stationary state in the limit
of large $N$~\cite{biroli_22}. However, the MFPT to a target for finite $N>1$ 
\blue{has not been computed for protocol B either}.

In this paper, we compute analytically the MFPT to the target by
$N$ Brownian searchers for both resetting protocols A and B defined
above. For the optimal reset rate $r^*$, we find a rather interesting
and somewhat surprising result for both protocols. 
We show that the MFPT, as a function of the reset rate $r$, exhibits
a unique minimum at $r=r^*$. However the optimal value $r^*$ is strictly
positive, i.e.,
the resetting is beneficial only for $N\le 7$ in protocol A and
$N\le 6$ in protocol B. When $N\ge 8$ in protocol A or $N\ge 7$ in protocol B,
\blue{the optimal reseting rate becomes $r^*=0$}. In those cases, the MFPT is a monotonically 
increasing function of $r$
with a minimum at $r=0$, implying that resetting will only increase
the mean search time and hence is detrimental to the search process. 
To understand the
origin of these two magic numbers $N=7$ and $N=6$ in the two protocols, 
it is convenient to continue analytically our general formula for integer $N$
to real $N$. Following the analytic continuation, we show that
the actual transitions take place respectively at $N_c=7.3264773\ldots$
(for protocol A) and $N_c=6.3555864\ldots$ (for protocol B) which
turn out to be the unique roots of two different transcendental equations.

The rest of the paper is organized as follows. \blue{In Section \ref{sec:MFPT}, we briefly recall how to compute the MFPT from the survival probability}. In Section \ref{subsection:A}
and Section \ref{subsection:B}, we present the exact computations
of the MFPT, respectively in protocol A and protocol B.
We conclude in Section \ref{sec:Conclusion} and some details
of the computations are presented in the Appendix.

\begin{figure}
\centering
\begin{minipage}[b]{0.49\textwidth}
\centering
\includegraphics[width = \textwidth]{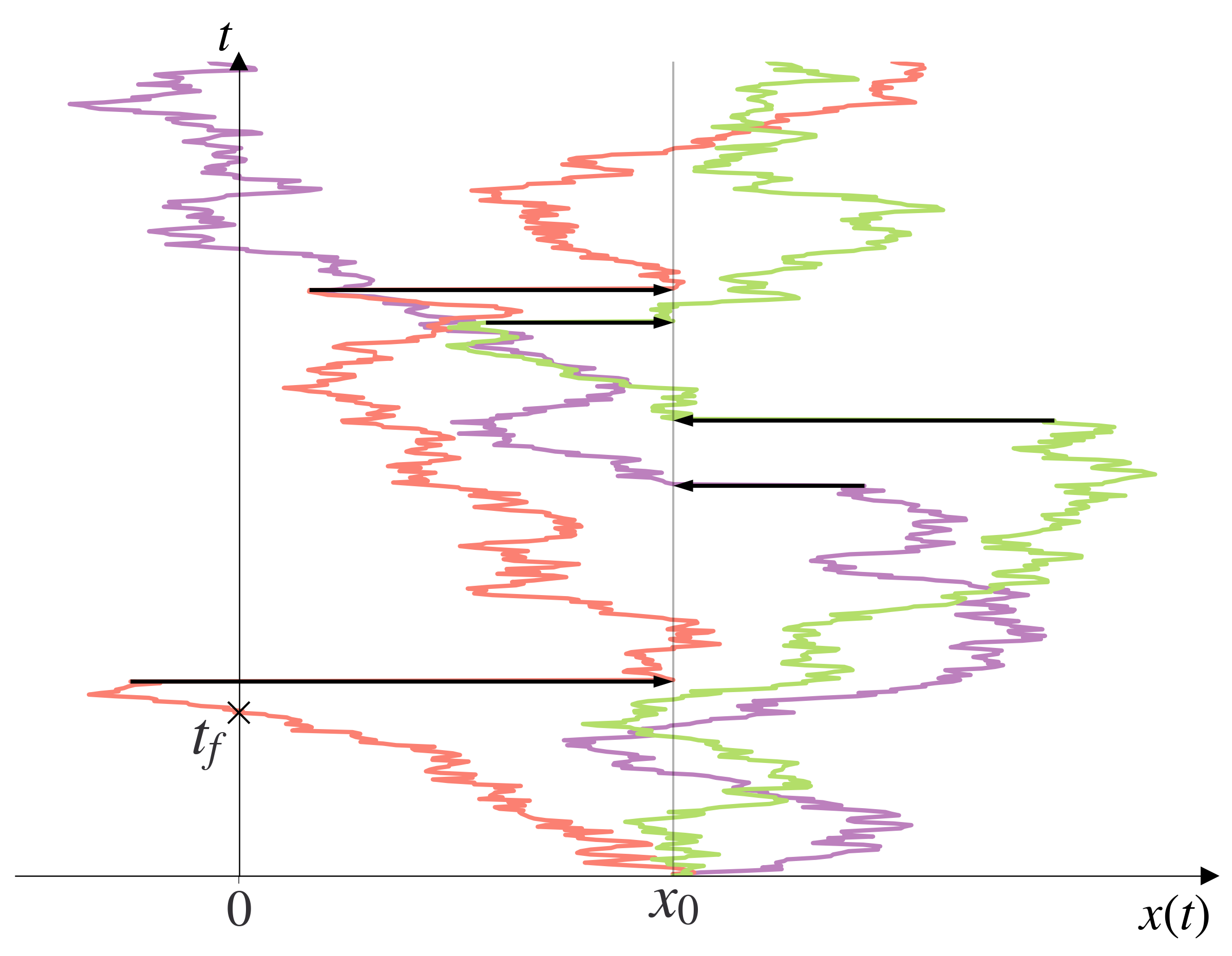}
%\caption{Protocol A}
\end{minipage}\hfill
\begin{minipage}[b]{0.49\textwidth}
\centering
\includegraphics[width = \textwidth]{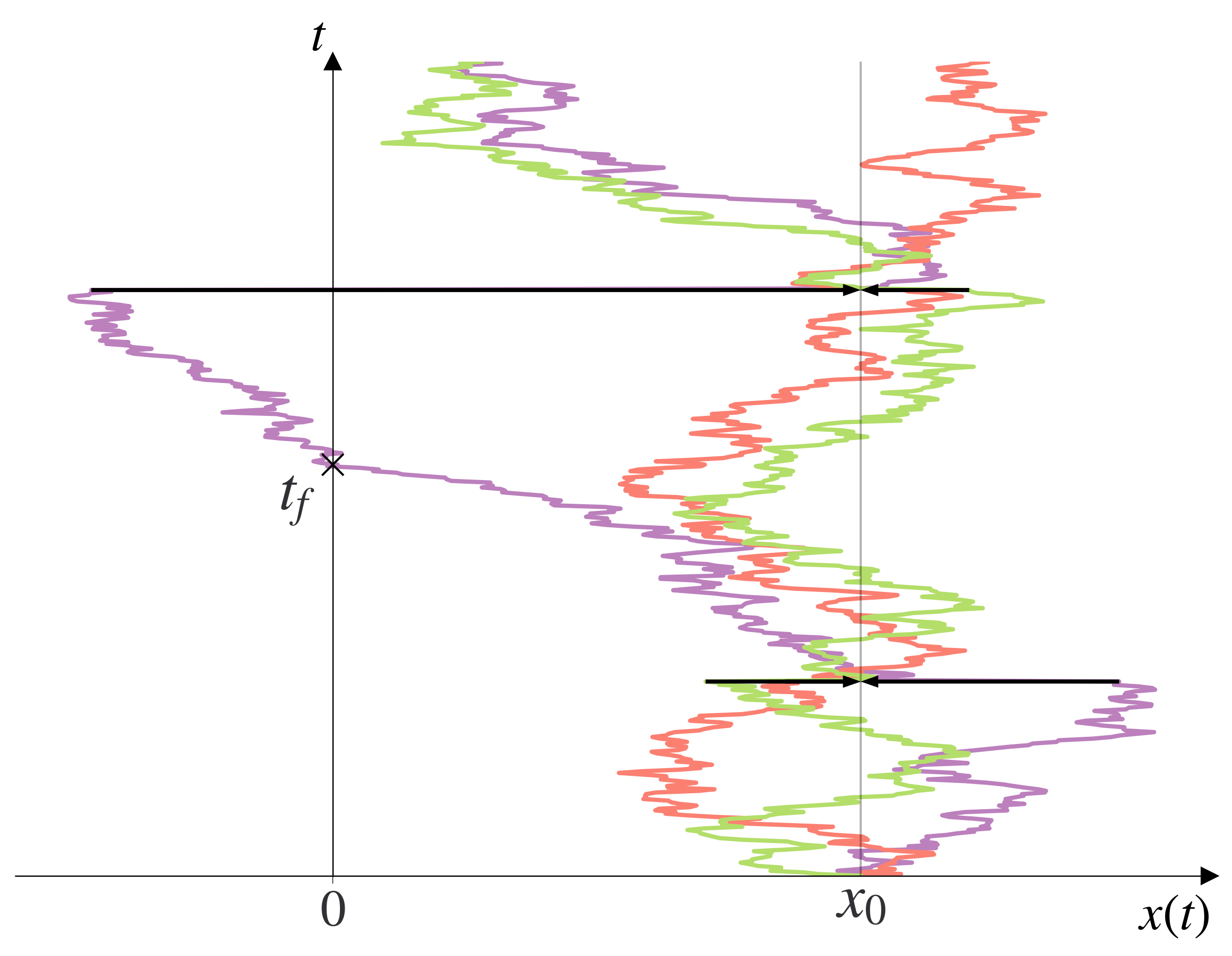}
%\caption{Protocol B}
\end{minipage}
\caption{Typical trajectories for $N = 3$ one-dimensional 
random walkers undergoing {\it independent} 
resetting (protocol A) in the left panel 
and {\it simultaneous} resetting (protocol B) in the right panel. 
Different colors correspond to different walkers and the resetting events 
are shown with full black arrows. The walkers all start at $x_0 > 0$ and 
reset to $x_0$ and $t_f$ denotes the first-passage time of the walkers to 
the target located at the origin $x = 0$.} 
\label{fig:sketch}
\end{figure}

\section{Mean first-passage time}
\label{sec:MFPT}

We consider $N$ Brownian particles that start at $x_0>0$ at $t=0$
and undergo stochastic resetting with rate $r$ following either the
protocol A or B defined above. We consider a stationary target
at the origin. Whenever any of the $N$ walkers reaches the origin,
the search is terminated. We denote by $t_f$ the first-passage time
to the origin by this $N$-particle process (see Fig. \ref{fig:sketch}). 
Clearly $t_f$ is a random variable and we will denote the MFPT by
$\langle t_f\rangle ^{(A)}_{r, N}(x_0)$ for protocol A and $\langle 
t_f\rangle^{(B)}_{r, N}(x_0)$ for protocol B. In order to compute $\langle 
t_f\rangle^{(A/B)}_{r, N}(x_0)$ it is useful to 
consider the cumulative distribution of $t_f$
\bea
S_{r, N}^{(A/B)}(x_0, t)= {\rm Prob.}\left[t_f\ge t\right]\, ,
\label{surv.1}
\eea
known as the survival probability, i.e., the probability
that none of the walkers have reached the target up to time $t$.
Using Eq. (\ref{surv.1}), the MFPT can then be expressed quite
generally for any process as~\cite{Redner_07,bf_05,Bray_13}
\bea 
\label{eq:def_Tx}
\langle t_f \rangle^{(A/B)}_{r, N}(x_0) = 
\int_0^{+\infty} t \left(-\pdv{S_{r, N}^{(A, B)}(x_0, t)}{t}\right) \dd t 
= \int_0^{+\infty} S_{r,N}^{(A/B)}(x_0, t) \dd t \;,
\eea
where, \blue{in the second equality}, we used integration by parts and assumed that 
$t\, S_{r, N}^{(A/B)}(x_0, t) \to 0$ when $t \to +\infty$, which
can be verified a posteriori. Hence to compute the MFPT we
need to compute the survival probability 
$S_{r, N}^{(A/B)}(x_0, t)$. We will now treat protocols
A and B separately.

\section{Protocol A}
\label{subsection:A}

In protocol A we have $N$ independent copies of a one-dimensional 
resetting random walker (see the left panel of Fig. \ref{fig:sketch}). These walkers are independent at all times $t$. 
Hence
\bea \label{eq:N21}
S_{r,N}^{(A)}(x_0, t) = \left[Q_{r}(x_0, t)\right]^N \;,
\eea
where $Q_r(x_0, t)$ is the survival probability of a single walker 
in the presence of resetting, starting at $x_0>0$ at $t = 0$.
This survival probability for a single resetting walker has been 
extensively studied \cite{evans_11, evans_11_2}. Let
us briefly recall the derivation here for the sake
of completeness. 

For a single walker, using a renewal approach, one can relate the resetting 
survival probability to 
the survival probability without resetting ($r = 0$), namely~\cite{evans_20}
\bea \label{eq:renewal}
Q_{r}(x_0, t) = e^{-r t} Q_{0}(x_0, t) + r \int_0^{+\infty} 
\dd \tau e^{-r \tau} Q_{0}(x_0, \tau) Q_{r}(x_0, t - \tau) \;.
\eea
This equation can be understood as follows. The first term
in Eq. (\ref{eq:renewal}) represents the probability of the event
when there are no resettings in the interval $[0,t]$ and the
particle survives up to $t$, starting at $x_0$. The probability
of no resetting in $[0,t]$ is 
$e^{-r\, t}$ and it then gets multiplied
by the probability $Q_0(x_0,t)$ that the particle survives 
without resetting up to $t$, leading to the first term in
Eq. (\ref{eq:renewal}).  
In the complementary case when the walker resets at least once to $x_0$, 
let us denote by $t - \tau$ 
the time of the last resetting event before $t$. Then, with probability $r\, \dd 
\tau$ the walker resets at $t - \tau$ and with probability $e^{-r \tau}$ 
the walker does not reset again in $[t - \tau,\, t]$. 
In the interval $[0,t-\tau]$ the survival probability is just
$Q_r(x_0,t-\tau)$, while in $[t-\tau,\, t]$ the survival probability
is $Q_0(x_0,\tau)$ since there is no resetting in $[t-\tau,t]$.
Using the renewal property of the process we then take the product
of all these probabilities
and integrate over all $\tau\in [0,t]$, leading to the second term
in Eq. (\ref{eq:renewal}).

The convolution structure of the renewal equation naturally calls for \blue{the use of} 
Laplace transform with respect $t$ defined as
\begin{equation}
\tilde{Q}_r(x_0,s)= \int_0^{\infty} Q_r(x_0,t)\, e^{-s\, t}\, dt\, .
\label{laplace.def}
\end{equation}
Taking the Laplace transform of Eq. (\ref{eq:renewal}) and simplifying
yields the result~\cite{evans_20}
\bea 
\label{eq:def_tQr}
\tilde{Q}_{r}(x_0, s) = 
\frac{\tilde{Q}_{0}(x_0, s + r)}{1 - r\, \tilde{Q}_{0}(x_0, s + r)} \;.
\eea
Furthermore, the survival probability of a standard one-dimensional Brownian motion \blue{without resetting} 
is given by the well known formula~\cite{Redner_07,bf_05,Bray_13}
\bea
\label{S1.4}
Q_{0}(x_0, t) = \erf\left( \frac{x_0}{\sqrt{4 D t}} \right) \;,
\eea
where ${\rm erf}(z)= (2/\sqrt{\pi})\, \int_0^z e^{-u^2}\, du$.
Its Laplace transform is given by
\bea 
\label{eq:def_tQ0}
\tilde{Q}_{0}(x_0, s) = \int_0^{+\infty} e^{- s t} 
\erf\left( \frac{x_0}{\sqrt{4 D t}} \right)  
\dd t = \frac{1}{s}\left( 1 - e^{- x_0 \sqrt{\frac{s}{D}}} \right) \;.
\eea

For simplicity, from now on, we re-write all the variables in terms of 
their dimensionless counterparts, i.e.,
\bea
\label{S1.5}
S = \frac{x_0^2}{D} s, \quad\, R = \frac{x_0^2}{D} r, 
\quad\, T = \frac{D}{x_0^2} t \;.
\eea
Inserting the result from Eq. (\ref{eq:def_tQ0}) in 
Eq. (\ref{eq:def_tQr}) gives, in terms of dimensionless variables,
\bea
\label{S1.7}
\tilde{Q}_{r}(x_0, s) = \frac{x_0^2}{D}\, \frac{1 - 
e^{-\sqrt{S + R}}}{\left[S + R e^{-\sqrt{S + R}}\right]} \;.
\eea
Inverting this Laplace transform formally one gets
\bea
\label{S1.8}
Q_{r}(x_0, t) = \int_\Gamma \frac{\dd S}{2\pi i} \; e^{S\, T} 
\frac{1 - e^{-\sqrt{S + R}}}{S + R \,e^{-\sqrt{S + R}}}\equiv q(R, T)\;.
\eea
where $\Gamma$ denotes the Bromwich contour in the complex $S$ plane.
Plugging this result in Eq. (\ref{eq:N21}) and then using 
Eq.~(\ref{eq:def_Tx}) we get
the dimensionless MFPT
\bea
\langle T_f \rangle^{(A)}(R, N) = \frac{D}{x_0^2} 
\langle t_f \rangle^{(A)}_{r, N}(x_0) = \int_0^{+\infty} 
\left[q(R, T)\right]^N\, \dd T \;. 
\label{eq:res_mfpt_A}
\eea

\begin{figure}
\centering
\begin{minipage}[b]{0.49\textwidth}
\centering
\includegraphics[width = \textwidth]{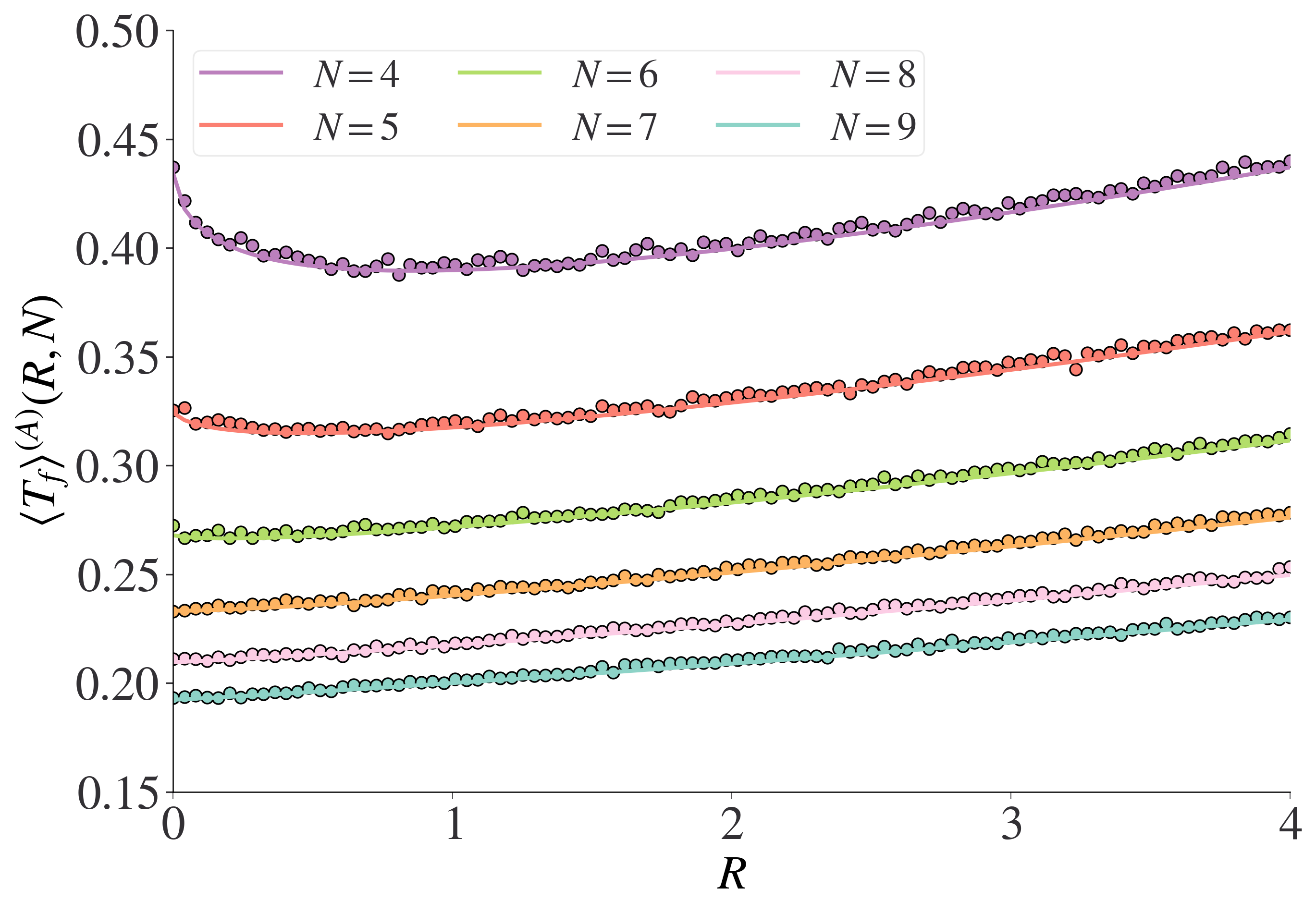}
%\caption{Protocol A}
\end{minipage}\hfill
\begin{minipage}[b]{0.49\textwidth}
\centering
\includegraphics[width = \textwidth]{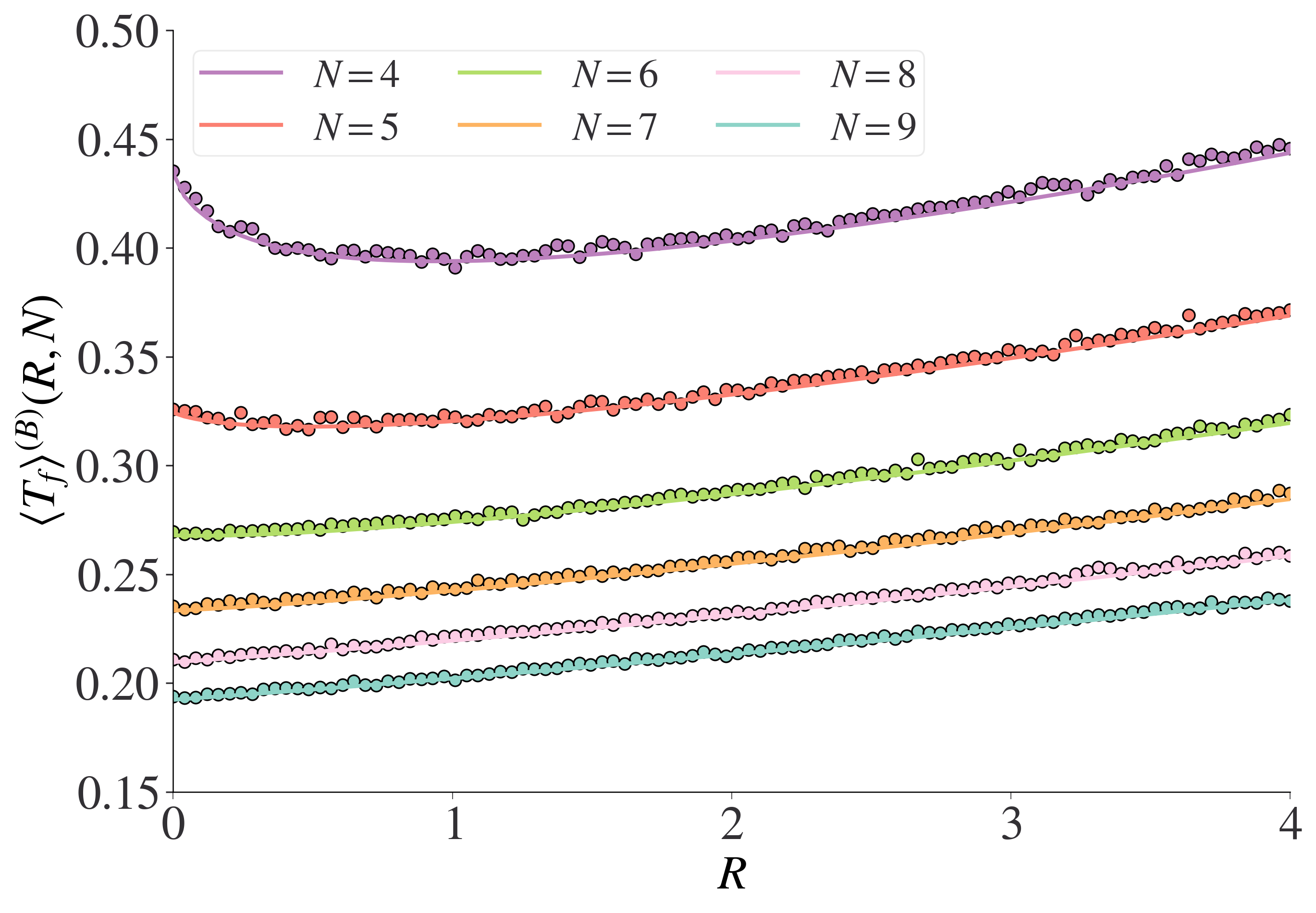}
%\caption{Protocol B}
\end{minipage}
\caption{
Comparison of theoretical and Monte Carlo results for the mean 
first-passage time as a function of the resetting rate for
protocol A (left panel) and protocol B 
(right panel). The solid lines correspond to the theoretical results 
given in Eq. (\ref{eq:res_mfpt_A}) (left panel) and Eqs. 
(\ref{eq:res_mfpt_B}) and (\ref{S2.5}) (right panel). 
The quantity $q(R,T)$ in Eq. (\ref{eq:res_mfpt_A}) is computed
by evaluating the Bromwich integral in Eq. (\ref{S1.8}) numerically.
The dots represent the results 
from Monte-Carlo simulations with $10^5$ samples. The different 
colors correspond to different values of $N$, where $N$ goes from 4 to 9 
from top to bottom. Notice that in both panels we can observe the 
gradual disappearance of the minimum at $R^\star > 0$.} 
\label{fig:mfpt}
\end{figure}

We inverted the Laplace transform in Eq. (\ref{S1.8}) numerically
and then evaluated the integral in Eq. (\ref{eq:res_mfpt_A}).
In the right panel of Fig.~\ref{fig:mfpt} we compare this theoretical prediction
with \blue{numerical Langevin} simulation results by plotting 
$\langle T_f \rangle^{(A)}(R, N)$
as a function of $R$, for different values of $N$.
We find excellent agreement.
Physically it is clear that as $R \to +\infty$ we 
expect the MFPT $\langle T_f \rangle^{(A)}(R, N)$ to diverge since the 
system constantly resets and thus never explores the space. This can be seen 
by noting that $q(R, T) \to 1$ as $R \to +\infty$ in Eq. (\ref{S1.8})
and hence the integral 
of the MFPT in Eq. (\ref{eq:res_mfpt_A}) diverges.
Let us now investigate the opposite limit $R\to 0$. 
If the MFPT decreases at small $R$ 
then it is likely 
that there is a certain $R^\star > 0$ where the
curve becomes a global minimum, before starting to increase again
and finally diverging as $R\to \infty$ (see Fig. \ref{fig:mfpt}). 
However if the MFPT increases for small $R$, then clearly $R^\star = 0$,
provided the MFPT increases monotonically with increasing $R$ as it
happens to be the case (see Fig. \ref{fig:mfpt}). 
Thus, the existence of a minimum $R^\star>0$ can then be investigated
by analyzing the small $R$ behavior of $\langle T_f \rangle^{(A)}(R, N)$. 

The small $R$ asymptotic 
behavior of $\langle T_f \rangle^{(A)}(R, N)$ depends on the value of 
$N$. It can be analyzed using Eqs. (\ref{eq:res_mfpt_A})
and (\ref{S1.8}), as shown in detail in the Appendix.
In fact, even though the search process makes sense only
for integer $N$, our analytical result in Eqs. (\ref{S1.8}) and (\ref{eq:res_mfpt_A}) can be continued analytically to real $N$.
Hence, from now on, we will consider $N$ real in this sense.
It turns out that
if $N \leq 2$ then $\langle T_f \rangle^{(A)}(R, N)$ diverges 
as $R \to 0$, if $2<N \leq 4$ then $\langle T_f \rangle^{(A)}(0, N)$
is finite but the slope of $\langle T_f 
\rangle^{(A)}(R, N)$ at $R \to 0$ is negatively divergent and 
finally if $N > 4$ 
both the MFPT and its derivative are finite at $R=0$. 
Let us summarize here the leading small $R$ behavior 
of the MFPT for different values of $N$:
\bea
\langle T_f \rangle^{(A)}(R, N) 
\blue{\underset{R \to 0}{\sim}}
%\stackrel{R \ll 1}{\sim}
\begin{dcases}
        \frac{\Gamma\left(1 - N/2\right)}{\pi^{N/2} N^{1 - N/2}}
\frac{1}{R^{1 - N/2}} \quad &\mbox{ if } N < 2 \;,\\
        -\frac{1}{\pi} \ln R \quad &\mbox{ if } N = 2 \;,\\
        \blue{C^{(A)}_N} -
\frac{2\, \Gamma(2-N/2)}{(N-2)\, \pi^{N/2}}\, N^{N/2-1}\, R^{N/2-1}
\quad &\mbox{ if } 2 < N < 4 \;, \\
       \blue{C^{(A)}_N}+ \frac{4}{\pi^2}
R \ln R \quad  &\mbox{ if } N = 4 \;,\\
        \blue{C^{(A)}_N} + R \int_0^{+\infty}
\left[q(0, T)\right]^{N-1} \pdv{q(R, T)}{R} \Big|_{R = 0} \dd
T \quad &\mbox{ if } N > 4 \;,
\end{dcases} 
\label{summary_A}
\eea
where for any $N > 2$ \blue{the constant $C^{(A)}_N$ is given by}
\bea
\blue{C_N^{(A)}} = \langle T_f \rangle^{(A)}(0, N) = \int_0^{+\infty}
\erf\left(\frac{1}{\sqrt{4 T}}\right)^N \dd T \;.
\label{value_A}
\eea

For $N\le 4$, the small $R$ behavior
of the MFPT above, 
combined with the divergence as $R\to \infty$,
indicates the existence of a finite $R^\star > 0$ for all $N\le 4$. 
However for $N > 4$ one has to find the condition
for a nonzero $R^\star > 0$.
For $N > 4$ both the MFPT and its first derivative with respect 
to $R$ are finite and the sign of the derivative can be
either positive or negative, depending on $N$. In fact, by taking the derivative
of Eq. (\ref{eq:res_mfpt_A}) and setting $R=0$ one gets
\bea \label{eq:AN4}
\pdv{\langle T_f \rangle^{(A)}(R, N)}{R} \Big|_{R = 0} = 
N\, \int_{0}^{+\infty} \left[q(0, T)\right]^{N-1} 
\pdv{q(R, T)}{R} \Big|_{R = 0} \dd T \;,
\eea
where, using Eq. (\ref{S1.4}), one has 
\bea
q(0,T)={\rm erf}\left(\frac{1}{\sqrt{4\, T}}\right)\, .
\label{q0T.1}
\eea
Taking the derivative of Eq. (\ref{S1.8}) with respect to $R$ and setting
$R=0$ gives
\bea
\pdv{q(R, T)}{R} \Big|_{R = 0}= \int_\Gamma \frac{\dd S}{2\pi i} \; e^{S\, T}\,
\left[\frac{1}{2\,S^{3/2}}\, e^{-\sqrt{S}}-\frac{1}{S^2}\, \left(
e^{-\sqrt{S}}- e^{-2\, \sqrt{S}}\right)\right]\, .
\label{der_lap.1}
\eea
This Laplace inversion can be explicitly done to give
\bea
\pdv{q(R, T)}{R} \Big|_{R = 0} = (T+1)\, \text{erf}\left(\frac{1}{ 
\sqrt{4\,T}}\right)-(T+2)\, \text{erf}\left(\frac{1}{\sqrt{T}}\right)+
\frac{2\,\sqrt{T}}{\sqrt{\pi}}\, 
\left( e^{-\frac{1}{4\, T}}- e^{-\frac{1}{T}}\right) +1 \;.
\label{S1.10}
\eea
Plugging Eqs. (\ref{q0T.1}) and (\ref{S1.10}) in Eq. (\ref{eq:AN4})
gives us the derivative of the MFPT at $R=0$ in terms of a single
integral, which unfortunately is not easy to evaluate
explicitly. However, it can be easily evaluated numerically for all
$N>4$ using Mathematica (see Fig. \ref{fig:dT}).
As $N$ increases beyond $4$, the derivative at $R=0$ in Eq. (\ref{eq:AN4})
increases, being negative initially, as can be seen in Fig. (\ref{fig:dT}).
As long as this derivative at $R=0$ is negative, we have a nonzero $R^*>0$.
When the derivative changes sign and becomes positive, we have $R^*=0$.
\blue{Using a dichotomous algorithm}, we find that this change of sign occurs at
$N_c =7.3264773 \cdots$. This is our main result in this section.
It says that the resetting in protocol A is beneficial for a
team of $N$ searchers as long as $N<N_c$. When $N>N_c$, resetting
increases the search time and hence is no longer a useful strategy.

\section{Protocol B}
\label{subsection:B}

In protocol B the {\it simultaneous} resetting (see the right panel of Fig. \ref{fig:sketch}) induces strong long range 
correlations between the walkers \cite{biroli_22}. Hence, the system is 
not simply $N$ independent copies of a single resetting random walker. 
However, since the resetting happens {\it simultaneously} we have a new 
renewal equation for the $N$-particle stochastic process
\bea
\label{S2.1}
S_{r, N}^{(B)}(x_0, t) = e^{-r t}\, S_{0, N}^{(B)}(x_0, t) + 
r \int_0^{+\infty} \dd \tau \; e^{-r \tau}\, S_{0, N}^{(B)}(x_0, \tau)\, 
S_{r, N}^{(B)}(x_0, t - \tau) \;.
\eea
The explanation of this renewal equation is exactly similar
to Eq. (\ref{eq:renewal}), except that one has to think in terms
of an $N$-particle process as a whole.
Now note that without resetting, i.e., for $r = 0$, the walkers become 
independent and hence using Eq. (\ref{S1.4}) we have
\bea \label{eq:Br0}
S_{0, N}^{(B)}(x_0, t) = \left[Q_0(x_0, t)\right]^N = 
\left[\erf\left( \frac{x_0}{\sqrt{4 D t}} \right)\right]^N \;.
\eea
As was done in Eq. (\ref{eq:renewal}), taking the Laplace transform of 
Eq. (\ref{S2.1}) we obtain
\bea
\tilde{S}_{r, N}^{(B)}(x_0, s) = \frac{\tilde{S}_{0, N}^{(B)}(x_0, 
s + r)}{1 - r \tilde{S}_{0, N}^{(B)}(x_0, s + r)} \;.
\label{S2.2}
\eea
Finally using Eq. (\ref{eq:def_Tx}) we can express the MFPT as
\bea
\label{S2.3}
\langle t_f \rangle^{(B)}_{r, N}(x_0) = 
\tilde{S}_{r, N}^{(B)}(x_0, s = 0) = 
\frac{\tilde{S}_{0, N}^{(B)}(x_0, r)}{1 - 
r \tilde{S}_{0, N}^{(B)}(x_0, r)} \;.
\eea
Inserting Eq. (\ref{eq:Br0}) in Eq. (\ref{S2.3})
we then get an explicit formula
\bea
\label{S2.4}
\langle t_f \rangle^{(B)}_{r, N}(x_0) = \frac{\int_0^{+\infty} \dd 
t\, e^{-r t}\, \left[\erf\left(\frac{x_0}{\sqrt{4 D t}} 
\right)\right]^N}{1 - r\, \int_0^{+\infty} \dd t\, e^{-r t}\, 
\left[\erf\left(\frac{x_0}{\sqrt{4 D t}} \right)\right]^N} \;.
\eea
Once again we appropriately re-scale the variables to make them 
dimensionless by setting $T = \frac{D}{x_0^2} t$ and $R = 
\frac{x_0^2}{D} r$ and obtain the simpler expression
\bea
\langle T_f \rangle^{(B)}(R, N) = \frac{D}{x_0^2} \langle 
t_f \rangle^{(B)}_{r, N}(x_0) = \frac{\int_0^{+\infty} \dd T\, 
e^{- R T}\, \left[\erf\left(\frac{1}{\sqrt{4 T}} \right)\right]^N}{1 - 
R \int_0^{+\infty} \dd T\, e^{-R T}\, \left[\erf\left(\frac{1}{\sqrt{4 T}} 
\right)\right]^N} = \frac{h(R, N)}{1 - R \, h(R, N)} \;, 
\label{eq:res_mfpt_B}
\eea
where for simplicity we introduced the function
\bea
\label{S2.5}
h(R, N) = \int_0^{+\infty} \dd T\, e^{- R T}\, \left[\erf\left(\frac{1}{\sqrt{4 T}} 
\right)\right]^N \;.
\eea
We verified this theoretical result
by comparing it to \blue{numerical} Langevin simulations as shown in the right panel of Fig. \ref{fig:mfpt}. As in the
case of protocol A, we infer the existence or not of a nonzero
optimal $R^*>0$ by analyzing the small $R$ behavior of Eq. (\ref{eq:res_mfpt_B}). 
The detailed derivation of the small $R$ behavior for different $N$ is given
in the Appendix. Here we summarize these results:
\bea
\hspace*{-0.4cm}\langle T_f \rangle^{(B)}(R, N) \stackrel{R \ll 1}{\sim}
\begin{dcases}
        \frac{\Gamma\left(1 - N/2\right)}{\pi^{N/2}}
\frac{1}{R^{1 - N/2}} \quad &\mbox{ if } N < 2\\
        -\frac{1}{\pi} \ln R \quad &\mbox{ if } N = 2\\
       C_N^{(B)} +
\frac{\Gamma\left(2 - N/2\right)(2 - N/2)}{\pi^{N/2} }
R^{N/2 - 1} \quad &\mbox{ if } 2 < N < 4\\
       C_N^{(B)}+ \frac{1}{\pi^2}\,
R\, \ln R \quad  &\mbox{ if } N = 4\\
       C_N^{(B)} +  \left\{ \left[
\int_0^{+\infty} \left[\erf\left( \frac{1}{\sqrt{4 T}} \right)\right]^N\,
\dd T \right]^2 - \int_0^{+\infty} T\,\left[\erf\left( \frac{1}{\sqrt{4 T}}
\right)\right]^N\, \dd T  \right\}\, R \quad &\mbox{ if } N > 4 \;,
\end{dcases}  \nonumber \\
\label{summary_B}
\eea
where for any $N > 2$, $C_N^{(B)}$ is a constant given by  
\bea
C_N^{(B)} = \langle T_f \rangle^{(B)}(0, N) = \langle T_f \rangle^{(A)}(0, N) =
\int_0^{+\infty} \left[\erf\left(\frac{1}{\sqrt{4 T}}\right)\right]^N \dd T \;.
\label{value_B}
\eea

\begin{figure}
\centering
\includegraphics[width = 0.6\textwidth]{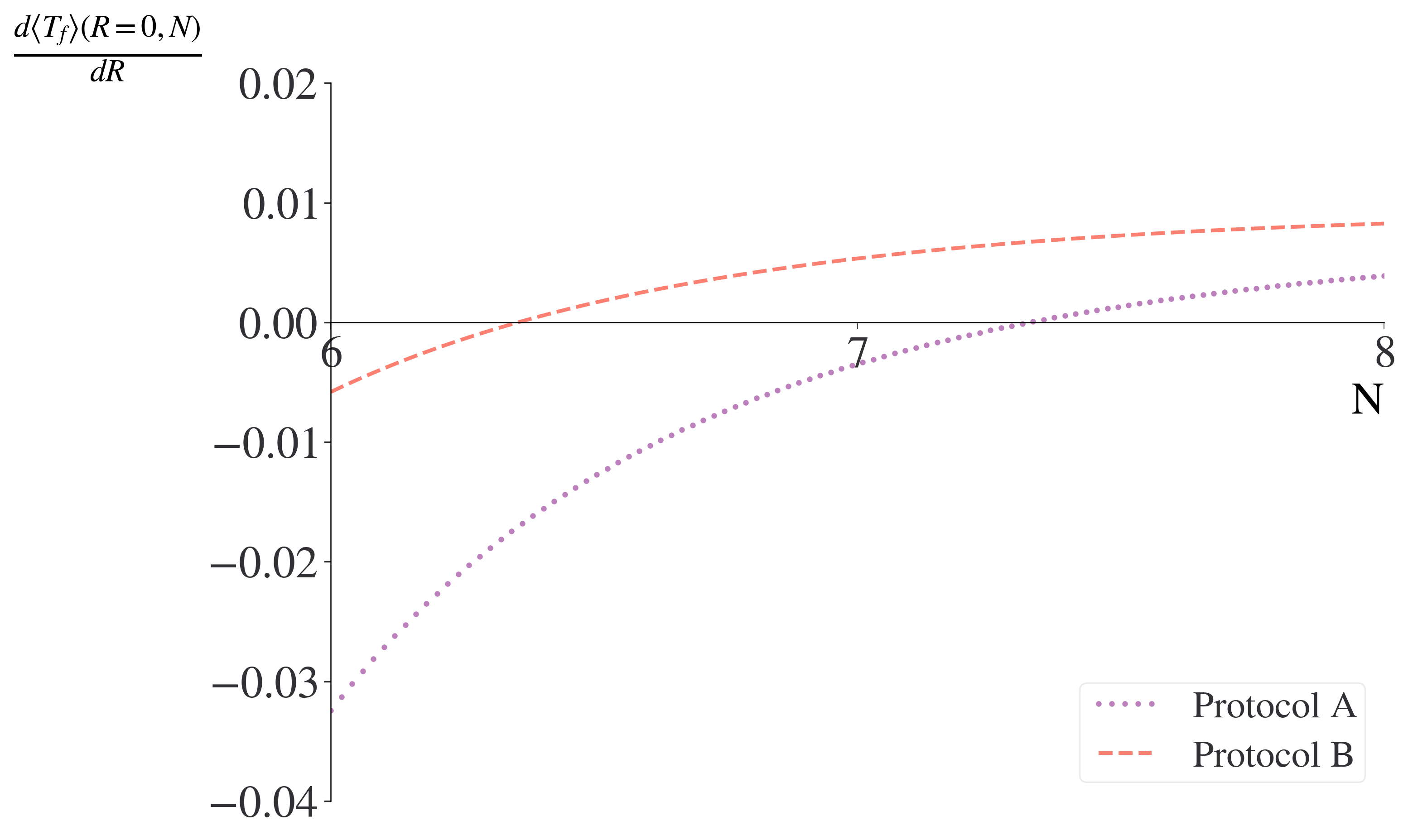}
\caption{The derivative of the MFPT at $R=0$ in Eq. (\ref{eq:AN4}) 
(protocol A) and Eq. (\ref{eq:res_dT_B})(protocol B), plotted
as a function of $N$ for $6<N<8$. The derivatives change sign
respectively at $N_c=7.3264773\ldots$
(protocol A) and $N_c=6.3555864\ldots$ (protocol~B).}
\centering
\label{fig:dT}
\end{figure}

As in the case of protocol A, it is clear from the small $R$ behavior
that there is an optimal $R^*>0$ for all $N\le 4$. For
$N>4$, both $h(R, N)$ and its first derivative are 
convergent when $R \to 0$. Then the derivative of the MFPT, for $N>4$, 
is given by
\begin{equation}
\pdv{\langle T_f \rangle^{(B)}(R, N)}{R} \Big|_{R = 0} = 
\left[h(0, N)\right]^2 + 
\partial_R h(R, N) \Big|_{R = 0} = \left[\int_0^{+\infty} 
\left[\erf\left( \frac{1}{\sqrt{4 T}} \right)\right]^N\, \dd T \right]^2 - 
\int_0^{+\infty} T\, \left[\erf\left( \frac{1}{\sqrt{4 T}} \right)\right]^N
\dd T \;. 
\label{eq:res_dT_B}
\end{equation}
The existence of a finite $R^\star > 0$ is uniquely determined by the 
sign of the above expression. If it is negative then there exists a 
finite $R^\star > 0$. However if it is positive then $R^\star = 0$ and 
the resetting hinders the search process. 
Once again, the integrals in Eq. (\ref{eq:res_dT_B}) can be
easily evaluated using Mathematica 
(see Fig. \ref{fig:dT}) and we find that the
critical value of $N$ defined as the value for which the derivative of the MFPT at $R = 0$ 
changes sign is given by $N_c = 6.3555864\ldots$. Thus, for protocol B, resetting benefits the search process
as long as $N<N_c$, but delays the search process for $N>N_c$.
The value of $N_c$ is slightly smaller in protocol B than that 
of the protocol A, reflecting the presence of an effective
attractive interaction between
the walkers in protocol B. 

\section{Conclusion}
\label{sec:Conclusion}

To summarize, in this paper we have studied analytically the mean 
first-passage time to a target at the origin in one dimension
by $N$ Brownian walkers all starting at $x_0>0$ and undergoing
diffusion with stochastic resetting. We considered two resetting
protocols: (A) where each walker diffuses and resets to $x_0$ with rate $r$
{\em independently} and (B) each walker diffuses independently
but resets {\em simultaneously} to $x_0$ with rate $r$. While in protocol
A, the walkers remain {\em uncorrelated} at all times, in protocol B
they become strongly {\em correlated} dynamically via simultaneous resetting.
We showed that in both protocols, the mean first-passage time,
as a function of the resetting rate $r$, has a minimum at $r=r^*>0$
as long as $N<N_c$, but for $N>N_c$ the optimal resetting rate is $r^*=0$.
The value of $N_c$ is slightly different in the two protocols.
Continuing our results analytically to real $N$, we showed that
$N_c=7.3264773\ldots$ for protocol A, while $N_c=6.3555864\ldots$ 
for protocol B. The main conclusion of our work is that
resetting is beneficial for the search process only when $N<N_c$.
For $N>N_c$, resetting hinders the search process. Our analytical
results have been verified in \blue{numerical} Langevin simulations.

The mean first-passage time for a single $N=1$ walker
has already been measured in optical tweezer experiments in
one~\cite{TPSRR_20,BBPMC_20} and two dimensions~\cite{FBPCM_21}.
It would be interesting to see if these measurements can be
extended to the $N>1$ case presented here, and in particular
to verify our theoretical predictions for $N_c$ in the two
protocols.

There are a number of other interesting directions in which our work may 
be extended. It would be interesting to find the critical values
$N_c$ in higher dimensions for both resetting protocols. 
Finally, one may investigate the mean first-passage time 
for interacting walkers and for non-diffusive processes such 
as L\'evy flights, using both resetting protocols.

\appendix

\section*{Appendix}
\label{App:A}

In this Appendix we provide a detailed
derivation of the small $R$ behavior of the scaled MFPT
$\langle T_f \rangle^{(A, B)}(R, N)$ for different
values of $0<N\le 4$. The results for protocols A and B 
are derived separately in the two following sections.

\section{Protocol A}
\label{App:A1}

For protocol A, the MFPT is given by Eq. (\ref{eq:res_mfpt_A})
that reads
\bea
\langle T_f \rangle^{(A)}(R, N) =\int_0^{+\infty}
\left[q(R, T)\right]^N\, \dd T \;,
\label{MFPT.A1}
\eea
where
\bea
q(R,T)= \int_\Gamma \frac{\dd S}{2\pi i} \; e^{S\, T}
\frac{1 - e^{-\sqrt{S + R}}}{S + R e^{-\sqrt{S + R}}}\, .
\label{qRT.A1}
\eea
In particular,
\bea
q(0,T)= {\rm erf}\left(\frac{1}{\sqrt{4\, T}}\right)\, .
\label{q0T.A1}
\eea
Now, if we put $R=0$ in Eq. (\ref{MFPT.A1}) and use Eq. (\ref{q0T.A1})
we get
\bea
\langle T_f \rangle^{(A)}(0, N)= \int_0^{\infty} \left[{\rm erf}
\left(\frac{1}{\sqrt{4\, T}}\right)\right]^N\, dT\, .
\label{R0.A1}
\eea
Using ${\rm erf}(z)\approx (2/\sqrt{\pi})\, z$ as $z\to 0$, one finds
that the integrand in Eq. (\ref{R0.A1}) behaves as $T^{-N/2}$ for large $T$.
Hence the integral is convergent for $N>2$, but diverges for $N\le 2$. 
This divergence for $N\le 2$ stems from the large $T$ behavior of the 
integrand. This indicates
that the behavior near $R=0$ depends crucially on $N$ and is delicate
to extract analytically. 
Since the divergence at $R=0$ comes from the large $T$ behavior
of the integrand for $N\le 2$, in order to extract the leading singular
behavior of the MFPT near $R=0$,   
it is necessary to investigate the scaling limit of $q(R,T)$ in 
Eq. (\ref{qRT.A1}) when
$R\to 0$, $T\to \infty$ while keeping the product $RT$ fixed.
One can then substitute this scaling form of $q(R,T)$ in Eq. (\ref{MFPT.A1})
and investigate the singular behavior of the MFPT as $R\to 0$.

To extract the scaling behavior of $q(R,T)$, we take the limit
$R\to 0$ and $S\to 0$ in Eq. (\ref{qRT.A1}), while keeping the
ratio $\tilde{S}=S/R$ fixed. Keeping $z=R\,T$ fixed, we get
to leading order for small $R$
\bea
q(R,T) \approx \frac{1}{\sqrt{R}}\, \int_\Gamma \frac{\dd \tilde{S}}{2\pi i} 
\; e^{\tilde{S}\, z}\, \frac{1}{\sqrt{1+\tilde{S}}}=  
\frac{1}{\sqrt{\pi\, R\, z}}\, e^{-z}\, .
\label{qRT_scaling.A1}
\eea
Consequently, in this scaling limit, we have
\begin{equation}
q(R,T) \approx \frac{1}{\sqrt{T}}\, f(R\,T)\, ; \quad\, {\rm where}\quad\,  
f(z)= \frac{1}{\sqrt{\pi}}\, e^{-z}\, .
\label{eq:def_f}
\end{equation}
Substituting this leading scaling behavior of $q(R,T)$ in
Eq. (\ref{MFPT.A1}), we then compute the small $R$ behavior
of the MFPT. Below we treat the five different cases
$N < 2$, $N = 2$, $2 < N < 4$, $N = 4$ and $N>4$ separately
in five subsections. 

\subsection{\blue{The case $N < 2$}}

In this case, substituting the scaling form of $q(R,T)$ from 
Eq. (\ref{eq:def_f}) in Eq. (\ref{MFPT.A1}), we get
\bea
\langle T_f \rangle^{(A)}(R, N) \approx \int_0^{+\infty} \left[ 
\frac{1}{\sqrt{T}} f(R T)\right]^N \dd T = \frac{R^{N/2 - 1}}{\pi^{N/2}} 
\int_0^{+\infty} e^{- N u} u^{-N/2} \dd u = \frac{\Gamma(1 - N/2)}{\pi^{N/2}}\, 
N^{N/2 - 1}\, R^{N/2 - 1} \;.
\label{A1.4}
\eea
Hence we see that, as long as $N < 2$, the integral converges
in Eq. (\ref{A1.4}) 
and the MFPT diverges as $\sim R^{N/2 - 1}$ as $R\to 0$.

\subsection{\blue{The case $N = 2$}}

In the $N = 2$ case we have to be a bit more careful. To start with, 
we split the integral in Eq. (\ref{MFPT.A1}) into three regions:
$T\in [0,1]$, $T\in [1,1/R]$ and $T\in [1/R,\infty)$. In the third part where
$T$ is large, we can approximate $q(R,T)$ by its scaling form in
Eq. (\ref{eq:def_f}). This gives
\bea
\langle T_f \rangle^{(A)}(R, N) \approx \int_0^{1} 
\left[q(R, T)\right]^2\, \dd T + \int_1^{1/R} 
\left[q(R, T)\right]^2\, \dd T + 
\int_{1/R}^{+\infty} \frac{1}{T}\, \left[f(R T)\right]^2  \dd T \;.
\label{A2.1}
\eea
Changing variable to $z = RT$ in the third integral, we see
that it is $O(1)$ since $f(z)=e^{-z}/\sqrt{\pi}$. Hence
\bea
\langle T_f \rangle^{(A)}(R, N) \approx \int_0^{1} 
\left[q(R, T)\right]^2\, \dd T + \int_1^{1/R}
\left[q(R, T)\right]^2\, \dd T  
+ \mathcal{O}(1) \;.
\label{A2.2}
\eea
For $T\ll 1/R$, 
the process is typically not resetting and hence we can replace
$q(R,T)\approx q(0,T)= {\rm erf}\left(1/\sqrt{4T}\right)$ in the
first two integrals. This gives 
\bea
\langle T_f \rangle^{(A)}(R, N) \approx  
\int_0^1 \dd T\, \left[{\rm erf}\left(\frac{1}{\sqrt{4T}}\right)\right]^2
+\int_1^{1/R} \dd T\, \left[{\rm erf}\left(\frac{1}{\sqrt{4T}}\right)\right]^2
+ \mathcal{O}(1) \;.
\label{A2.3}
\eea
The first integral is clearly $\mathcal{O}(1)$ and the principal
divergence comes from the second integral which is dominated
by the integrand near the upper limit $1/R$. Since $T>1$, we can
now expand ${\rm erf}\left(1/\sqrt{4T}\right)$ as a power series
in $1/\sqrt{T}$. The first term gives ${\rm erf}\left(1/\sqrt{4T}\right)
\approx 1/\sqrt{\pi\, T}$. Substituting this behavior
in the second integral in Eq. (\ref{A2.3}) gives the
leading order divergence
\bea
\langle T_f \rangle^{(A)}(R, N) \approx \frac{1}{\pi} 
\int_{1}^{1/R} \frac{dT}{T} +\mathcal{O}(1)
= - \frac{1}{\pi} \ln R + \mathcal{O}(1) \;.
\label{A2.5}
\eea

\subsection{\blue{The case $2 < N < 4$}}

In this case the integral in Eq. (\ref{MFPT.A1}) is convergent for $R=0$ and 
is given by Eq. (\ref{R0.A1}).
However the sub-leading term turns out to be singular as $R\to 0$.
To derive the subleading term, it is useful to analyze
the derivative at $R=0$.
Indeed, deriving 
Eq. (\ref{MFPT.A1}) with respect to $R$ gives 
\bea
\label{eq:derA}
\pdv{\langle T_f \rangle^{(A)}(R, N)}{R}=  N\, \int_0^{+\infty} 
\left[q(R,T)\right]^{N-1}\, \pdv{q(R, T)}{R}\, \dd T  \; .
\eea
Now we replace $q(R,T)$ by its scaling form in Eq. (\ref{eq:def_f})
and make the change of variable $z=RT$. This gives
\bea
\pdv{\langle T_f \rangle^{(A)}(R, N)}{R} 
\approx  N\, R^{N/2-2}\, \int_0^{+\infty}
z^{1-N/2}\, \left[f(z)\right]^{N-1}\, f'(z)\, \dd z\, . 
\label{A3.2}
\eea
Using $f(z)= e^{-z}/\sqrt{\pi}$ and performing the integral
exactly gives
\bea
\pdv{\langle T_f \rangle^{(A)}(R, N)}{R}\approx 
-\frac{\Gamma(2 - N/2)}{\pi^{N/2}}\, \Gamma(2-N/2)\, R^{N/2 - 2}\; .
\label{A3.3}
\eea
Note that this is well defined for $N < 4$, otherwise the 
Gamma is diverging. Integrating it back with respect to $R$
gives the small $R$ asymptotic behavior of the MFPT
\begin{equation}
\langle T_f \rangle^{(A)}(R, N)\approx \int_0^{\infty} 
\left[{\rm erf}\left(\frac{1}{\sqrt{4T}}\right)\right]^{N}\, \dd T
- \frac{2\, \Gamma(2-N/2)}{(N-2)\, \pi^{N/2}}\, N^{N/2-1}\, R^{N/2-1} \, .
\label{A3.4} 
\end{equation}
Clearly as $R$ increases from $0$, the MFPT decreases due to the
negative sign of the second term in Eq. (\ref{A3.4}), indicating
that the minimum of the MFPT occurs at $R^*>0$.

\subsection{\blue{The case $N = 4$}}

In the $N = 4$ case, the analysis is somewhat similar
to the $N=2$ case. In this case, the derivative in Eq. (\ref{eq:derA})
reads
\bea
\pdv{\langle T_f \rangle^{(A)}(R, 4)}{R}=  4\, \int_0^{+\infty}
\left[q(R,T)\right]^{3}\, \pdv{q(R, T)}{R}\, \dd T  \; .
\label{A4.1}
\eea
We anticipate in this case, and verify a posteriori, that
as in the case $N=2$, the scaling form of $q(R,T)$ only
gives a $\mathcal{O}(1)$ contribution and the leading
divergence as $R\to 0$ in Eq. (\ref{A4.1}) has a different source.
So, we need to go beyond the scaling regime and estimate
both $q(R,T)$ and $\partial_R q(R,T)$ for small $R$.
The first one is simple since we already know explicitly that
$q(0,T)= {\rm erf}\left(1/\sqrt{4T}\right)$. To estimate the
derivative for small $R$, we note that
for $T \ll 1/R$, the diffusing particle typically hardly resets and
hence
\bea
q(R, T) \approx e^{- R\, T} q(0, T)\approx q(0,T)- R\, T\, q(0,T)\, .
\label{A4.2}
\eea
Taking a derivative with respect to $R$ gives the estimate
\bea
\pdv{q(R, T)}{R} \approx - T\, e^{-R\, T}\, q(0, T) \sim - T\, q(0, T)\;.
\label{A4.3}
\eea

To proceed, we now split the integral in Eq. (\ref{A4.1})
into three regimes: $[0,1]$, $[1,1/R]$ and $[1/R,\infty)$,
\begin{equation}
\pdv{\langle T_f \rangle^{(A)}(R, 4)}{R}=4\, \int_0^{1} \left[q(R,T)\right]^{3}\, \pdv{q(R, T)}{R}\, \dd T 
+4\, \int_{1}^{1/R} 
\left[q(R,T)\right]^{3}\, \pdv{q(R, T)}{R}\, \dd T  
+4\, \int_{1/R}^{\infty} 
\left[q(R,T)\right]^{3}\, \pdv{q(R, T)}{R}\, \dd T  \, .
\label{A4.31}
\end{equation}
In the third integral, denoted by $I_3$, we can use the scaling form of 
$q(R,T)$ in Eq. (\ref{eq:def_f}) and get, after the customary change of
variables $z=RT$,
\bea
I_3 \approx 4\, \int_{1}^{\infty} \frac{\dd z}{z}\, \left[f(z)\right]^3\, f'(z)
\sim \mathcal{O}(1) \, ,
\label{A4.32}
\eea
where we used $f(z)= e^{-z}/\sqrt{\pi}$.
In the first two integrals, in contrast, we cannot use the scaling form.
Instead, we can replace $q(R,T)$ and $\partial_R q(R,T)$ by
their approximate forms in Eqs. (\ref{A4.2}) and (\ref{A4.3})
respectively. It is easy to check that after this substitution,
the first integral $I_1$ over $[0,1]$ in Eq. (\ref{A4.31}) 
is $\mathcal{O}(1)$.
Hence the leading divergence in Eq. (\ref{A4.31}) comes
from the second integral $I_2$ over $[1,1/R]$,
which then reads
\begin{equation}
I_2\approx 4\, \int_{1}^{1/R} \left[q(0,T\right]^3\, 
\left(- T\, q(0,T)\right)\, \dd T
= -4\, \int_1^{1/R} T\, \left[{\rm erf}\left(\frac{1}{\sqrt{4T}}\right)
\right]^4\, \dd T\, .
\label{A4.33}
\end{equation}
In this range,
since $T>1$, we can again expand ${\rm erf}\left(1/\sqrt{4T}\right)$
in a Taylor series in powers of $1/\sqrt{T}$. The first term in
this expansion provides the leading divergence, and we get
\begin{equation}
I_2\approx \frac{4}{\pi^2}\, \int_1^{1/R} \frac{dT}{T}\approx 
\frac{4}{\pi^2}\, \ln R\, .
\label{A4.34}
\end{equation}
Adding the three integrals, we then find that as $R\to 0$
\bea
\pdv{\langle T_f \rangle^{(A)}(R, 4)}{R} \approx  
\frac{4}{\pi^2}\, \ln R + \mathcal{O}(1)\, .
\label{A4.35}
\eea
Integrating back with respect to $R$, we then get the leading small
$R$ behavior of the MFPT
\bea
\langle T_f \rangle^{(A)}(R, 4)\approx \int_0^{\infty}
\left[{\rm erf}\left(\frac{1}{\sqrt{4T}}\right)\right]^{4}\, \dd T
+ \frac{4}{\pi^2}\, R\, \ln R \,.
\label{A4.36}
\eea
Note that the sub-leading term is negative for small $R$, indicating
that the MFPT decreases from its $R=0$ value as $R$ increases.
This again implies that the MFPT has a nonzero minimum at some $R^*>0$.

\subsection{\blue{The case $N>4$}}

Finally, in the fifth case when $N>4$, both the MFPT and its
first derivative are finite at $R=0$. Hence, the sub-leading
behavior for $N>4$ is linear as $R\to 0$. As discussed
in the main text, the sign of the sub-leading linear
term changes from \blue{negative to positive} as $N$ crosses $N_c=7.7.3264773\ldots$
from below.

\vspace{0.5cm}

The different behaviors of $\langle T_f \rangle^{(A)}(R, N) $ for small $R$ are summarized in Eq. (\ref{summary_A}) in the text. 

%\blue{\bf GS: Since we already give the summary in Eq. (\ref{summary_A}) we could simply refer to that equation instead of repeating it -- the Appendices are already rather long compared to the size of the paper !?}
%
%We conclude this section
%by summarizing the small $R$ asymptotics of the MFPT 
%in protocol A in a single formula as follows:
%\bea
%\langle T_f \rangle^{(A)}(R, N) \stackrel{R \ll 1}{\sim}
%\begin{dcases}
%        \frac{\Gamma\left(1 - N/2\right)}{\pi^{N/2} N^{1 - N/2}}
%\frac{1}{R^{1 - N/2}} \quad &\mbox{ if } N < 2\\
%        -\frac{1}{\pi} \ln R \quad &\mbox{ if } N = 2\\
%        \langle T_f \rangle^{(A)}(0, N) -
%\frac{2\, \Gamma(2-N/2)}{(N-2)\, \pi^{N/2}}\, N^{N/2-1}\, R^{N/2-1}
%\quad &\mbox{ if } 2 < N < 4\\
%        \langle T_f \rangle^{(A)}(0, N) + \frac{4}{\pi^2}
%R \ln R \quad  &\mbox{ if } N = 4\\
%        \langle T_f \rangle^{(A)}(0, N) + R \int_0^{+\infty}
%\left[q(0, T)\right]^{N-1} \pdv{q(R, T)}{R} \Big|_{R = 0} \dd
%T \quad &\mbox{ if } N > 4
%\end{dcases} \;,
%\label{summary_A1}
%\eea
%where for any $N > 2$ 
%\bea
%\langle T_f \rangle^{(A)}(0, N) = \int_0^{+\infty}
%\left[\erf\left(\frac{1}{\sqrt{4 T}}\right)\right]^N \dd T \;.
%\eea

\section{Protocol B}
\label{App:B1}

\blue{For protocol B, we recall that the MFPT is given by Eq. (\ref{eq:res_mfpt_B}), namely} 
\begin{equation}
\langle T_f \rangle^{(B)}(R, N) =\frac{h(R, N)}{1 - R \, h(R, N)} \;,
\label{MFPT.B1}
\end{equation}
where the function $h(R,N)$, given in Eq. (\ref{S2.5}), reads
\begin{equation}
h(R, N) = \int_0^{+\infty} \dd T\, e^{- R T}\, 
\left[\erf\left(\frac{1}{\sqrt{4 T}}
\right)\right]^N \;.
\label{def_hRN.B1}
\end{equation}
It is convenient to make a change
of variable $u=R\,T$ in Eq. (\ref{def_hRN.B1}) and re-write it as
\bea
h(R, N) = \frac{1}{R} \int_0^{+\infty} \dd u \; e^{-u}\,
\left[\erf\left(\sqrt{\frac{R}{4 u}}\right)\right]^N \;.
\label{B1.1}
\eea
Putting directly $R=0$ in Eq. (\ref{MFPT.B1})
gives the same result as in Eq. (\ref{R0.A1}) for protocol A, namely
\begin{equation}
\langle T_f \rangle^{(B)}(0, N)= h(0,N)= \int_0^{\infty} \left[{\rm erf}
\left(\frac{1}{\sqrt{4\, T}}\right)\right]^N\, \dd T\, .
\label{R0.B1}
\end{equation}
Again this integral is convergent only for $N>2$.

For later purposes, we will also need the first derivative of $h(R,N)$
with respect to $R$, which reads from Eq. (\ref{def_hRN.B1})
\bea
\pdv{h(R, N)}{R} =- \int_0^{+\infty} \dd T\, T\, e^{-R T}\, 
\left[\erf\left(\frac{1}{\sqrt{4 T}}\right)\right]^N \;.
\label{B3.1}
\eea
Performing the same change of variable $u = RT$, one obtains
an alternative expression
\bea
\pdv{h(R, N)}{R} = -\frac{1}{R^2} \int_0^{+\infty} \dd u \; u\, 
e^{-u}\, \left[\erf\left(\sqrt{\frac{R}{4 u}}\right)\right]^N \;.
\label{B3.2}
\eea

Our goal is to extract the asymptotic small $R$ behavior of $h(R,N)$
in Eq. (\ref{def_hRN.B1}) or equivalently in Eq. (\ref{B1.1}) 
for different $N$ and then use these
results in Eq. (\ref{MFPT.B1}) to derive the small $R$ behavior
of the MFPT. As in protocol A, we consider the five cases
$N < 2$, $N = 2$, $2 < N < 4$, $N = 4$ and $N>4$ separately
in the five subsections below.

\subsection{\blue{The case $N < 2$}}

When $N < 2$ the integral $h(R,N)$ in Eq. (\ref{B1.1}) 
becomes divergent as $R \to 0$ and the divergence ensues
from the large $u$ regime of the integrand. 
To compute this small $R$ divergence, we 
use $\erf(z) 
\approx (2/\sqrt{\pi})\, z$ for small $z$ in
Eq. (\ref{B1.1}) and carry out the integral.
This gives, to leading order as $R\to 0$,
\bea
h(R, N)\approx \frac{R^{N/2 - 1}}{\pi^{N/2}} 
\int_0^{+\infty} e^{-u}\, u^{-N/2}\, \dd u = \frac{\Gamma(1 - N/2)}{\pi^{N/2}} 
\, R^{N/2 - 1}\; .
\label{B1.3}
\eea
Note that this result 
is valid only for $N < 2$, as otherwise the Gamma function 
becomes divergent. Substituting this behavior of $h(R,N)$
in Eq. (\ref{MFPT.B1}) we get, to leading order for small $R$,
\bea
\langle T_f \rangle^{(B)}(R, N) \approx
\frac{\Gamma(1 - N/2)}{\pi^{N/2}}\, \frac{1}{R^{1 - N/2}} \;.
\label{B1.4}
\eea
This divergence of the MFPT as $R\to 0$, along with its divergence
as $R\to \infty$, indicates that the minimum of the MFPT
occurs at a nonzero $R^*>0$ for $N<2$.

\subsection{\blue{The case $N = 2$}}

For $N = 2$ we need to make a finer analysis of $h(R, N)$. In this case 
we split the integral in Eq. (\ref{B1.1}) into two regimes:
$u < R \ll 1$ and $u > R$. This  
leads to
\bea
h(R, 2) = \frac{1}{R} \int_0^R \dd u \; e^{-u} 
\left[\erf\left(\sqrt{\frac{R}{4 u}}\right)\right]^2  + 
\frac{1}{R} \int_R^{+\infty} \dd u \; e^{-u} 
\left[\erf\left(\sqrt{\frac{R}{4 u}}\right)\right]^2 \;.
\label{B2.1}
\eea
Then in the integrand in the first term, to leading order,
the ${\rm erf}$ function can be replaced by $1$, and 
hence the integral remains $\mathcal{O}(1)$ as $R\to 0$.
The divergence comes from the second integral, where
we can use $\erf(z)\approx (2/\sqrt{\pi})\, z$ for small $z$.
This gives
\bea
h(R, N)\approx \frac{1}{\pi} \int_R^{+\infty} \frac{\dd u}{u} \; e^{-u} 
+\mathcal{O}(1) \;.
\label{B2.2}
\eea
Integrating by parts, one immediately finds the
leading order behavior for small $R$,
\bea
h(R, N) \approx - \frac{1}{\pi}\, \ln R + \mathcal{O}(1)\;.
\label{B2.4}
\eea
Finally, substituting this in Eq. (\ref{MFPT.B1}), we get
\bea
\langle T_f \rangle^{(B)}(R, 2) \approx - \frac{1}{\pi}\, 
\ln R + \mathcal{O}(1)\;.
\label{B2.5}
\eea
Hence, the MFPT diverges logarithmically as $R\to 0$,
indicating that for $N=2$, we will again have
a nonzero $R^*$.

\subsection{Third case, $2 < N < 4$}

For $N > 2$, putting $R=0$ in Eq. (\ref{def_hRN.B1}),
one finds that $h(0,N)$ is finite and is given by Eq. (\ref{R0.B1}).
Hence $\langle T_f \rangle^{(B)}(0, N) = h(0, N) < +\infty$
from Eq. (\ref{MFPT.B1}). To extract the dominant
sub-leading term, it is convenient to first find
how the derivative of $h(R,N)$ diverges as $R\to 0$ 
by analyzing Eq. (\ref{B3.1}) or equivalently Eq. (\ref{B3.2}).
We insert the asymptotic small $z$ behavior
$\erf(z)
\approx (2/\sqrt{\pi})\, z$ in Eq. (\ref{B3.2})
to get the leading small $R$ behavior
\bea
\pdv{h(R, N)}{R} \approx - \frac{R^{N/2-2}}{\pi^{N/2}}\,
\int_0^{+\infty} \dd u \; e^{-u}\, u^{1 - N/2} = 
- \frac{\Gamma(2 - N/2)}{\pi^{N/2}}\, R^{N/2-2}\;.
\label{B3.3}
\eea
Note that the Gamma function is well defined for $N<4$.
Integrating it back with respect to $R$, we then get, 
up to the first sub-leading term,
\bea
h(R, N)\approx 
h(0, N) - \frac{2\,\Gamma(2 - N/2)}{(N-2)\,\pi^{N/2}}\, R^{N/2-1} \;.
\label{B3.4}
\eea
Finally, substituting this result for $h(R,N)$ in Eq. (\ref{MFPT.B1})
we get, noting that $(N/2-1)<1$, the following result
\bea
\langle T_f \rangle^{(B)}(R, N) \approx h(R,N)\approx
h(0, N) - \frac{2\,\Gamma(2 - N/2)}{(N-2)\,\pi^{N/2}}\, R^{N/2-1} \;.
\label{B3.5}
\eea
Note that the sub-leading term is negative for $2<N<4$, indicating
that the MFPT decreases from its value at $R=0$ as $R$ increases.
This again implies that the optimal $R^*>0$.

\subsection{\blue{The case, $N = 4$}}
In this case $h(0,4)$ in Eq. (\ref{def_hRN.B1}) is finite.
To extract the subleading behavior as $R\to 0$, we again analyze
the derivative in Eq. (\ref{B3.2}) by splitting
the integral into two regimes
$u<R\ll 1$ and $u>R$
\bea
\pdv{h(R, 4)}{R} = -\frac{1}{R^2}\, \int_0^{R} \dd u \; u\, e^{-u}\, 
\left[\erf\left(\sqrt{\frac{R}{4 u}}\right)\right]^4 -\frac{1}{R^2}\, 
\int_R^{+\infty} \dd u \; u\, e^{-u}\, \left[\erf\left(\sqrt{\frac{R}{4 u}}
\right)\right]^4 \; .
\label{B4.1}
\eea
We can replace the ${\rm erf}$ by $1$ in the integrand in the first term,
leading to an $\mathcal{O}(1)$ result for the first integral as $R\to 0$.
In the second integral, we use the small $z$ behavior of the error function 
$\erf(z)\approx (2/\sqrt{\pi})\, z$, which then gives
\bea
\pdv{h(R, 4)}{R} \approx \mathcal{O}(1)-\frac{1}{\pi^2}\, \int_R^{\infty}
\frac{du}{u}\, e^{-u}\, .
\label{B4.2}
\eea
Integrating by parts, one gets the
leading order behavior for small $R$,
\bea
\pdv{h(R, 4)}{R} \approx \frac{1}{\pi^2}\, \ln R \;.
\label{B4.3}
\eea
Integrating it back with respect to $R$ gives
\bea
h(R,4) \approx h(0,4) +\frac{1}{\pi^2}\, R\, \ln R \, .
\label{B4.4}
\eea
Finally, substituting this result in Eq. (\ref{MFPT.B1}) gives
the small $R$ asymptotics of the MFPT
\bea
\langle T_f \rangle^{(B)}(R, \blue{4}) 
\approx h(0,4)+\frac{1}{\pi^2}\, R\, \ln R \, .
\label{B4.5}
\eea
Since the second term is negative as $R\to 0$, we again see that
the MFPT decreases from its value at $R=0$ as $R$ increases,
indicating that the optimal $R^*>0$.

\subsection{\blue{The case $N>4$}}

Finally, in the fifth case when $N>4$, both $h(0,N)$ in 
Eq. (\ref{def_hRN.B1}) and 
its first derivative $h'(0,N)$ in Eq. (\ref{B3.1}) are finite. 
Expanding Eq. (\ref{MFPT.B1}) up to $\mathcal{O}(R)$, we then
get as $R\to 0$
\bea
\langle T_f \rangle^{(B)}(R, N) &\approx &
h(0,N) +\left(h(0,N+h'(0,N)\right)\, R \nonumber \\
&= &
\int_0^{\infty} \left[{\rm erf}
\left(\frac{1}{\sqrt{4\, T}}\right)\right]^N\, \dd T\,
+ \left\{ \left[
\int_0^{+\infty} \left[\erf\left( \frac{1}{\sqrt{4 T}} \right)\right]^N
\dd T \right]^2 - \int_0^{+\infty} T \left[\erf\left( \frac{1}{\sqrt{4 T}}
\right)\right]^N \dd T  \right\}\, R \;.\, \nonumber \\
\label{B5.1}
\eea
Hence, the sub-leading
behavior of the MFPT for $N>4$ is linear as $R\to 0$, as in protocol A.
The sign of the subleading linear term is negative for $N<N_c$
and is positive for $N>N_c$ where $N_c=6.3555864\ldots$.

%\blue{\bf GS: Same remark as above: since we already give the summary in Eq. (\ref{summary_B}) we could simply refer to that equation instead of repeating it -- the Appendices are already rather long compared to the size of the paper !?}

\vspace{0.5cm}
The different behaviors of  $\langle T_f \rangle^{(B)}(R, N)$ for small $R$ are summarized in Eq. (\ref{summary_B}) in the text.

%Finally, it is useful to summarize the small $R$ asymptotics
%of the MFPT for different $N$ in protocol B
%in a single formula.
%\bea
%\langle T_f \rangle^{(B)}(R, N) \stackrel{R \ll 1}{\sim}
%\begin{dcases}
%        \frac{\Gamma\left(1 - N/2\right)}{\pi^{N/2}} 
%\frac{1}{R^{1 - N/2}} \quad &\mbox{ if } N < 2\\
%        -\frac{1}{\pi} \ln R \quad &\mbox{ if } N = 2\\
%        \langle T_f \rangle^{(B)}(0, N) +
%\frac{\Gamma\left(2 - N/2\right)(2 - N/2)}{\pi^{N/2} } 
%R^{N/2 - 1} \quad &\mbox{ if } 2 < N < 4\\
%        \langle T_f \rangle^{(B)}(0, N) + \frac{1}{\pi^2}\, 
%R\, \ln R \quad  &\mbox{ if } N = 4\\
%        \langle T_f \rangle^{(B)}(0, N) +  \left\{ \left[ 
%\int_0^{+\infty} \left[\erf\left( \frac{1}{\sqrt{4 T}} \right)\right]^N 
%\dd T \right]^2 - \int_0^{+\infty} T \left[\erf\left( \frac{1}{\sqrt{4 T}} 
%\right)\right]^N \dd T  \right\}\, R \quad &\mbox{ if } N > 4
%\end{dcases} \;,
%\label{summary_B1}
%\eea
%where for any $N > 2$ we have that
%\bea
%\langle T_f \rangle^{(B)}(0, N) = \langle T_f \rangle^{(A)}(0, N) = 
%\int_0^{+\infty} \left[\erf\left(\frac{1}{\sqrt{4 T}}\right)\right]^N \dd T \;.
%\eea


\begin{thebibliography}{26}

\bibitem{bell}  W. J. Bell, {\it Searching Behaviour: The Behavioural Ecology of Finding Resources} (London:
Chapman and Hall, 1991)

\bibitem{adam} G. Adam, M. Delbr\"uck, {\it Reduction of dimensionality in 
biological diffusion processes: Structural Chemistry and Molecular 
Biology}, Eds A. Rich and N. Davidson (London: WH Freeman and Company, 
1968).


\bibitem{Metzler_14}
R. Metzler, S. Redner, G. Oshanin, {\it First-passage phenomena
and their applications} (World Scientific),  {\bf 35} (2014).


\bibitem{bartumeus} F. Bartumeus, J. Catalan,
J. Phys. A: Math. Theor. {\bf 42}, 434002 (2009).

\bibitem{viswanathan} G. M. Viswanathan, M. G. E. da Luz, 
E. P. Raposo, H. E. Stanley
{\it The Physics of Foraging: An Introduction to 
Random Searches and Biological Encounters} (Cambridge: 
Cambridge University Press, 2011).

\bibitem{berg} O. G. Berg, R. B. Winter, P. H. von Hippel,
Biochemistry {\bf 20}, 6929 (1981).

\bibitem{coppey} M. Coppey, O. B\'enichou, M. Moreau,  
Biophys. J. {\bf 87}, 1640 (2004).

\bibitem{ghosh} S. Ghosh, B. Mishra, A. B. Kolomeisky, D. Chowdhury, J. Stat. Mech. 123209, (2018).

\bibitem{chowdhury} D. Chowdhury, Biophys. J. {\bf 116}, 2057 (2019).

\bibitem{benichou_05} O. B\'enichou, M. Coppey, M. Moreau, P.-H. Suet,
R. Voituriez, Phys. Rev. Lett. {\bf 94}, 198101 (2005).

\bibitem{benichou_07} O. B\'enichou, M. Moreau, P.-H. Suet, R. Voituriez, 
J. Chem. Phys. {\bf 126}, 234109 (2007).

\bibitem{benichou_11} O. B\'enichou, C. Loverdo, M. Moreau, R. Voituriez,
Rev. Mod. Phys. {\bf 83}, 81 (2011).

\bibitem{villen} M. Villen-Altramirano, J. Villen-Altramirano, RESTART: A method for accelerating rare event simulations, in {\it Queueing, Performance and Control in ATM,} edited by J. W. Cohen and C. D.
Pack (North-Holland, Amsterdam, 1991).

\bibitem{luby} M. Luby, A. Sinclair, D. Zuckerman, 
Inf. Proc. Lett. {\bf 47}, 173 (1993).

\bibitem{tong} H. Tong, C. Faloutsos, J.-Y. Pan, Knowl. Inf. Syst. {\bf 14}, 327 (2008).

\bibitem{lorenz} J. H. Lorenz, {\it Runtime distributions and criteria for restarts}, in {\it SOFSEM 2018: Theory and Practice of Computer Science,} edited by A Tjoa, L. Bellatreche, S. Biffl, J. van Leeuwen, and
J. Wiedermann, Lecture Notes in Computer Science Vol. 10706 (Springer, Berlin, 2018).

\bibitem{evans_20} M. R. Evans, S. N. Majumdar, G. Schehr, 
J. Phys. A: Math. Theor. {\bf 53}, 193001 (2020).

\bibitem{evans_11} M. R. Evans, S. N. Majumdar, 
Phys. Rev. Lett. {\bf 106}, 160601 (2011).

\bibitem{evans_11_2} M. R. Evans, S. N. Majumdar, J. Phys. A: Math. Theor. 
{\bf 44}, 435001 (2011).

\bibitem{EM_14}
M. R. Evans, S. N. Majumdar, J. Phys. A: Math. Theor.  {\bf 47}, 285001 (2014).

\bibitem{KMSS_14}
L. Kusmierz, S. N. Majumdar, S. Sabhapandit, G. Schehr, Phys. Rev. Lett. {\bf 113}, 220602 (2014).

\bibitem{MSS_15}
S. N. Majumdar, S. Sabhapandit, G. Schehr, Phys. Rev. E {\bf 91}, 052131 (2015).

\bibitem{PKE_16}
A. Pal, A. Kundu, M. R. Evans, J. Phys. A: Math. Theor. {\bf 49}, 225001 (2016).

\bibitem{Reu_16}
S. Reuveni, Phys. Rev. Lett. {\bf 116}, 170601 (2016).

\bibitem{MV_16}
M. Montero, J. Villarroel, Phys. Rev. E {\bf 94}, 032132 (2016).

\bibitem{PR_17}
A. Pal, S. Reuveni, Phys. Rev. Lett. {\bf 118}, 030603 (2017).

\bibitem{BEM_17}
D. Boyer, M. R. Evans, S. N. Majumdar, J. Stat. Mech. 023208, (2017).
%Long time scaling behaviour for diffusion with resetting and memory.

\bibitem{CS_18}
A. Chechkin, I. M. Sokolov, Phys. Rev. Lett. {\bf 121}, 050601 (2018).

\bibitem{Bres_20}
P. C. Bressloff, J. Phys. A: Math. Theor. {\bf 53}, 425001 (2020).

\bibitem{Pinsky_20}
R. G. Pinsky, Stoch. Proc. Appl. {\bf 130}, 2954 (2020).
% Diffusive search with spatially dependent resetting


\bibitem{BMS_22}
B. De Bruyne, S. N. Majumdar, G. Schehr, Phys. Rev. Lett. {\bf 128}, 
200603 (2022).

\bibitem{TPSRR_20}
O. Tal-Friedman, A. Pal, A. Sekhon, S. Reuveni, Y. Roichman, 
J. Phys. Chem. Lett. {\bf 11}, 7350 (2020).


\bibitem{BBPMC_20}
B. Besga, A. Bovon, A. Petrosyan, S. N. Majumdar, S. Ciliberto, 
Phys. Rev. Research {\bf 2}, 032029(R) (2020).

\bibitem{FBPCM_21}
F. Faisant, B. Besga, A. Petrosyan, S. Ciliberto, S. N. Majumdar, 
J. Stat. Mech., 113203 (2021).

%Optimal Resetting Brownian Bridges via Enhanced Fluctuations

\bibitem{CS15} C. Christou, A. Schadschneider, J. Phys. A: Math. Theor.
{\bf 48}, 285003 (2015).

\bibitem{CM15} D. Campos, V. M\'endez, Phys. Rev. E {\bf 92}, 
062115 (2015).

\bibitem{Bel_18} S. Belan, Phys. Rev. Lett. {\bf 120}, 080601 (2018).

\bibitem{RMS19} S. Ray, D. Mondal, S. Reuveni, J. Phys. A: Math.
Theor. {\bf 52}, 255002 (2019).

\bibitem{ANBBD19} S. Ahmad, I. Nayak, A. Bansal, A. Nandi, D. Das,
Phys. Rev. E {\bf 99}, 022130 (2019).

\bibitem{Pal_19}
A. Pal, V. V. Prasad, Phy. Rev. Research {\bf 1}, 032001 (2019).

\bibitem{Pal_19_2}
A. Pal, V. V. Prasad, Phys. Rev. E. {\bf 99}, 032123 (2019).

\bibitem{Vasquez_20}
G. Mercado-Vasquez, D. Boyer, S. N. Majumdar, G. Schehr, 
J. Stat. Mech., 113203 (2020).

\bibitem{Vasquez_22}
G. Mercado-Vasquez, D. Boyer, S. N. Majumdar, J. Stat. Mech., 063203 (2022).

\bibitem{Vasquez_22_2}
G. Mercado-Vasquez, D. Boyer, S. N. Majumdar, J. Stat. Mech., 093202 (2022).


\bibitem{VAM22} O. Vilk, M. Assaf, B. Meerson,
Phys. Rev. E {\bf 106}, 024117 (2022).

\bibitem{biroli_22} M. Biroli, H. Larralde, S. N. Majumdar, G. Schehr, arXiv preprint: 2211.00563
%{\it Extreme statistics and spacing distribution in a Brownian gas 
%correlated by resetting}, 

\bibitem{Redner_07}
S. Redner, {\it A Guide to First-Passage Processes} (Cambridge University Press, Cambridge, UK, 2007). 

\bibitem{Bray_13}
A. J. Bray, S. N. Majumdar, G. Schehr, Adv. Phys. {\bf 62}, 225 (2013).

\bibitem{bf_05} S. N. Majumdar, Curr. Sci. {\bf 89}, 2076 (2005).










\end{thebibliography}
\end{document}